\documentclass[9pt,twocolumn,twoside,]{pinp}

\providecommand{\tightlist}{%
  \setlength{\itemsep}{0pt}\setlength{\parskip}{0pt}}

\usepackage[T1]{fontenc}
\usepackage[utf8]{inputenc}

\definecolor{pinpblue}{HTML}{185FAF}  
\definecolor{pnasbluetext}{RGB}{0,101,165} 

\title{A Brief Introduction to Redis}

\author[1]{Dirk Eddelbuettel}

  \affil[1]{Department of Statistics, University of Illinois,
Urbana-Champaign, IL, USA}

\leadauthor{Eddelbuettel}

\begin{abstract}
This note provides a brief introduction to Redis highlighting its
usefulness in multi-lingual statistical computing.
\end{abstract}

\dates{This version was compiled on \today}

\doifooter{\url{https://cran.r-project.org/package=RcppRedis}}

\pinpfootercontents{Redis Intro}

\begin{document}

\verticaladjustment{-2pt}

\maketitle
\thispagestyle{firststyle}
\ifthenelse{\boolean{shortarticle}}{\ifthenelse{\boolean{singlecolumn}}{\abscontentformatted}{\abscontent}}{}


\hypertarget{overview-and-introduction}{%
\section{Overview and Introduction}\label{overview-and-introduction}}

\href{https://redis.io}{Redis} \citep{Redis} is a very popular, very
powerful, and very widely-used `in-memory database-structure store'. It
runs as a background process (a ``daemon'' in Unix lingo) and can be
accessed locally or across a network making it very popular choice for
`data caches'. There is more to say about \href{https://redis.io}{Redis}
than we possibly could in a \emph{short} vignette introducing it, and
other places already do so. The
\href{https://www.openmymind.net/2012/1/23/The-Little-Redis-Book/}{Little
Redis Book} \citep{Seguin:2021:Redis}, while a decade old (!!) is a
fabulous \emph{short} start. The \href{https://redis.io}{official site}
is very good as well (but by now a little overwhelming as so many
features have been added).

This vignette aims highlight two aspects: how easy it is to use
\href{https://redis.io}{Redis} on simple examples, and also to stress
how \href{https://redis.io}{Redis} enables easy \emph{multi-lingual}
computing as it can act as a `middle-man' between any set of languages
capable of speaking the \href{https://redis.io}{Redis} protocol -- which
may cover most if not all common languages one may want to use!

\hypertarget{data-structure-example-one-key-value-setter-and-getter}{%
\section{Data Structure Example One: Key-Value Setter and
Getter}\label{data-structure-example-one-key-value-setter-and-getter}}

We will show several simple examples for the

\begin{itemize}
\tightlist
\item
  \texttt{redis-cli} command used directly or via shell scripts
\item
  \textsf{Python} via the standard \textsf{Python} package for
  \href{https://redis.io}{Redis}
\item
  \textsf{C} / \textsf{C++} via the
  \href{https://github.com/redis/hiredis}{hiredis} library
\item
  \textsf{R} via \textbf{RcppRedis} \citep{CRAN:RcppRedis} utilising the
  \href{https://github.com/redis/hiredis}{hiredis} library
\end{itemize}

to demonstrate how different languages all can write to and read from
\href{https://redis.io}{Redis}. Our first example will use the simplest
possibly data structure, a simple \texttt{SET} and \texttt{GET} of a
key-value pair.

\hypertarget{command-line}{%
\subsection{Command-Line}\label{command-line}}

\texttt{redis-cli} is command-line client. Use is straightforward as
shown an simply key-value getter and setter. We show use in `shell
script' form here, but the same commands also work interactively.

\begin{Shaded}
\begin{Highlighting}[]
\CommentTok{\#\# note that document processing will show all}
\CommentTok{\#\# three results at once as opposed to one at time}
\ExtensionTok{redis{-}cli}\NormalTok{ SET ice{-}cream chocolate}
\ExtensionTok{redis{-}cli}\NormalTok{ GET ice{-}cream}
\ExtensionTok{redis{-}cli}\NormalTok{ GET ice{-}cream}
\CommentTok{\#  OK}
\CommentTok{\#  chocolate}
\CommentTok{\#  chocolate}
\end{Highlighting}
\end{Shaded}

Here, as in general, we will omit hostname and authentication arguments:
on the same machine, \texttt{redis-cli} and the background
\texttt{redis} process should work `as is'. For access across a (local
or remote) network, the configuration will have to be altered to permit
access at given network interfaces and IP address ranges.

We show the \href{https://redis.io}{redis} commands used in uppercase
notation, this is in line with the documentation. Note, however, that
\href{https://redis.io}{Redis} itself is case-insensitive here so
\texttt{set} is equivalent to \texttt{SET}.

\hypertarget{python}{%
\subsection{Python}\label{python}}

\href{https://redis.io}{Redis} does have bindings for most, if not all,
languages to computing with data. Here is a simple \textsf{Python}
example.

\begin{Shaded}
\begin{Highlighting}[]
\ImportTok{import}\NormalTok{ redis}

\NormalTok{redishost }\OperatorTok{=} \StringTok{"localhost"}
\NormalTok{redisserver }\OperatorTok{=}\NormalTok{ redis.StrictRedis(redishost)}

\NormalTok{key }\OperatorTok{=} \StringTok{"ice{-}cream"}
\NormalTok{val }\OperatorTok{=} \StringTok{"strawberry"}
\NormalTok{res }\OperatorTok{=}\NormalTok{ redisserver.}\BuiltInTok{set}\NormalTok{(key, val)}
\BuiltInTok{print}\NormalTok{(}\StringTok{"Set"}\NormalTok{, val, }\StringTok{"under"}\NormalTok{, key, }\StringTok{"with result"}\NormalTok{, res)}
\CommentTok{\#  Set strawberry under ice{-}cream with result True}
\NormalTok{key }\OperatorTok{=} \StringTok{"ice{-}cream"}
\NormalTok{val }\OperatorTok{=}\NormalTok{ redisserver.get(key)}
\BuiltInTok{print}\NormalTok{(}\StringTok{"Got"}\NormalTok{, val, }\StringTok{"from"}\NormalTok{, key)}
\CommentTok{\#  Got b\textquotesingle{}strawberry\textquotesingle{} from ice{-}cream}
\end{Highlighting}
\end{Shaded}

For \textsf{Python}, the redis commands are generally mapped to
(lower-case named) member functions of the instantiated redis connection
object, here \texttt{redisserver}.

\hypertarget{c-c}{%
\subsection{C / C++}\label{c-c}}

Compiled languages work similarly. For \textsf{C} and \textsf{C++}, the
\href{https://github.com/redis/hiredis}{hiredis} `minimalistic' library
from the \href{https://redis.io}{Redis} project can be used---as it is
by \textbf{RcppRedis}. Here we only show the code without executing it.
This example is included in the package are as the preceding ones.
\textsf{C} and \textsf{C++} work similarly to the interactive or
\textsf{Python} commands. A simplified (and incomplete, see the
\texttt{examples/} directory of the package for more) example of writing
to \href{https://redis.io}{Redis} would be

\begin{Shaded}
\begin{Highlighting}[]
\NormalTok{redisContext }\OperatorTok{*}\NormalTok{prc}\OperatorTok{;}    \CommentTok{// pointer to redis context}

\BuiltInTok{std::}\NormalTok{string}\OperatorTok{ }\NormalTok{host }\OperatorTok{=} \StringTok{"127.0.0.1"}\OperatorTok{;}
\DataTypeTok{int}\NormalTok{ port }\OperatorTok{=} \DecValTok{6379}\OperatorTok{;}

\NormalTok{prc }\OperatorTok{=}\NormalTok{ redisConnect}\OperatorTok{(}\NormalTok{host}\OperatorTok{.}\NormalTok{c\_str}\OperatorTok{(),}\NormalTok{ port}\OperatorTok{);}
\CommentTok{// should check error here}
\NormalTok{redisReply }\OperatorTok{*}\NormalTok{reply }\OperatorTok{=} \OperatorTok{(}\NormalTok{redisReply}\OperatorTok{*)}
\NormalTok{    redisCommand}\OperatorTok{(}\NormalTok{prc}\OperatorTok{,} \StringTok{"SET ice{-}cream }\SpecialCharTok{\%s}\StringTok{"}\OperatorTok{,}\NormalTok{ value}\OperatorTok{);}
\CommentTok{// should check reply here}
\end{Highlighting}
\end{Shaded}

Reading is done by submitting for example a \texttt{GET} command for a
key after which the \texttt{redisReply} contains the reply string.

\hypertarget{r}{%
\subsection{R}\label{r}}

The \textbf{RcppRedis} packages uses \textbf{Rcpp} Modules along with
\textbf{Rcpp} \citep{CRAN:Rcpp,TAS:Rcpp} to connect the
\href{https://github.com/redis/hiredis}{hiredis} library to \textsf{R}.
A simple \textsf{R} example follows.

\begin{Shaded}
\begin{Highlighting}[]
\FunctionTok{library}\NormalTok{(RcppRedis)}
\NormalTok{redis }\OtherTok{\textless{}{-}} \FunctionTok{new}\NormalTok{(Redis, }\StringTok{"localhost"}\NormalTok{)}
\NormalTok{redis}\SpecialCharTok{$}\FunctionTok{set}\NormalTok{(}\StringTok{"ice{-}cream"}\NormalTok{, }\StringTok{"hazelnut"}\NormalTok{)}
\CommentTok{\#  [1] "OK"}
\NormalTok{redis}\SpecialCharTok{$}\FunctionTok{get}\NormalTok{(}\StringTok{"ice{-}cream"}\NormalTok{)}
\CommentTok{\#  [1] "hazelnut"}
\end{Highlighting}
\end{Shaded}

\hypertarget{data-structure-example-two-hashes}{%
\section{Data Structure Example Two:
Hashes}\label{data-structure-example-two-hashes}}

\href{https://redis.io}{Redis} has support for a number of standard data
structures including hashes. The official documentation list
\href{https://redis.io/commands\#hash}{sixteen commands} in the
corresponding group covering writing (\texttt{hset}), reading
(\texttt{hget}), checking for key (\texttt{hexists}), deleting a key
(\texttt{hdel}) and more.

\begin{Shaded}
\begin{Highlighting}[]
\ExtensionTok{redis{-}cli}\NormalTok{ HSET myhash abc 42}
\ExtensionTok{redis{-}cli}\NormalTok{ HSET myhash def }\StringTok{"some text"}
\CommentTok{\#  1}
\CommentTok{\#  1}
\end{Highlighting}
\end{Shaded}

We can illustrate reading and checking from Python:

\begin{Shaded}
\begin{Highlighting}[]
\BuiltInTok{print}\NormalTok{(redisserver.hget(}\StringTok{"myhash"}\NormalTok{, }\StringTok{"abc"}\NormalTok{))}
\CommentTok{\#  b\textquotesingle{}42\textquotesingle{}}
\BuiltInTok{print}\NormalTok{(redisserver.hget(}\StringTok{"myhash"}\NormalTok{, }\StringTok{"def"}\NormalTok{))}
\CommentTok{\#  b\textquotesingle{}some text\textquotesingle{}}
\BuiltInTok{print}\NormalTok{(redisserver.hexists(}\StringTok{"myhash"}\NormalTok{, }\StringTok{"xyz"}\NormalTok{))}
\CommentTok{\#  False}
\end{Highlighting}
\end{Shaded}

For historical reasons, \textbf{RcppRedis} converts to/from \textsf{R}
internal serialization on hash operations so it cannot \emph{directly}
interoperate with these examples as they not deploy \textsf{R}-internal
representation. The package has however a `catch-all' command
\texttt{exec} (which excutes a given \textsf{Redis} command string)
which can be used here:

\begin{Shaded}
\begin{Highlighting}[]
\NormalTok{redis}\SpecialCharTok{$}\FunctionTok{exec}\NormalTok{(}\StringTok{"HGET myhash abc"}\NormalTok{)}
\CommentTok{\#  [1] "42"}
\NormalTok{redis}\SpecialCharTok{$}\FunctionTok{exec}\NormalTok{(}\StringTok{"HGET myhash def"}\NormalTok{)}
\CommentTok{\#  [1] "some text"}
\NormalTok{redis}\SpecialCharTok{$}\FunctionTok{exec}\NormalTok{(}\StringTok{"HEXISTS myhash xyz"}\NormalTok{)}
\CommentTok{\#  [1] 0}
\end{Highlighting}
\end{Shaded}

\hypertarget{data-structure-example-three-sets}{%
\section{Data Structure Example Three:
Sets}\label{data-structure-example-three-sets}}

Sets are another basic data structure that is well-understood. In sets,
elements can be added (once), removed (if present), and sets can be
joined, intersected and differenced.

\begin{Shaded}
\begin{Highlighting}[]
\ExtensionTok{redis{-}cli}\NormalTok{ SADD myset puppy}
\ExtensionTok{redis{-}cli}\NormalTok{ SADD myset kitten}
\ExtensionTok{redis{-}cli}\NormalTok{ SADD otherset birdie}
\ExtensionTok{redis{-}cli}\NormalTok{ SADD otherset kitten}
\ExtensionTok{redis{-}cli}\NormalTok{ SINTER myset otherset}
\CommentTok{\#  1}
\CommentTok{\#  1}
\CommentTok{\#  1}
\CommentTok{\#  1}
\CommentTok{\#  kitten}
\end{Highlighting}
\end{Shaded}

We can do the same in \textsf{Python}. Here we showi only the final
intersect command---the set-addition commands are equivalent across
implementations as are most other \href{https://redis.io}{Redis}
command.

\begin{Shaded}
\begin{Highlighting}[]
\BuiltInTok{print}\NormalTok{(redisserver.sinter(}\StringTok{"myset"}\NormalTok{, }\StringTok{"otherset"}\NormalTok{))}
\CommentTok{\#  \{b\textquotesingle{}kitten\textquotesingle{}\}}
\end{Highlighting}
\end{Shaded}

And similarly in \textsf{R} where \texttt{exec} returns a list:

\begin{Shaded}
\begin{Highlighting}[]
\NormalTok{redis}\SpecialCharTok{$}\FunctionTok{exec}\NormalTok{(}\StringTok{"SINTER myset otherset"}\NormalTok{)}
\CommentTok{\#  [[1]]}
\CommentTok{\#  [1] "kitten"}
\end{Highlighting}
\end{Shaded}

\hypertarget{data-structure-example-four-lists}{%
\section{Data Structure Example Four:
Lists}\label{data-structure-example-four-lists}}

So far we have looked at setters and getters for values, hashes, and
sets. All of these covered only one value per key. But
\textsf{Redis} has support for many more types.

\begin{Shaded}
\begin{Highlighting}[]
\ExtensionTok{redis{-}cli}\NormalTok{ LPUSH mylist chocolate}
\ExtensionTok{redis{-}cli}\NormalTok{ LPUSH mylist strawberry vanilla}
\ExtensionTok{redis{-}cli}\NormalTok{ LLEN mylist}
\CommentTok{\#  1}
\CommentTok{\#  3}
\CommentTok{\#  3}
\end{Highlighting}
\end{Shaded}

We can access the list in \textsf{Python}. Here we show access by index.
Note that the index is zero-based, so `one' accesses the middle element
in a list of length three.

\begin{Shaded}
\begin{Highlighting}[]
\BuiltInTok{print}\NormalTok{(redisserver.lindex(}\StringTok{"mylist"}\NormalTok{, }\DecValTok{1}\NormalTok{))}
\CommentTok{\#  b\textquotesingle{}strawberry\textquotesingle{}}
\end{Highlighting}
\end{Shaded}

In \textsf{R}, using the `list range' command for elements 0 to 1:

\begin{Shaded}
\begin{Highlighting}[]
\NormalTok{redis}\SpecialCharTok{$}\FunctionTok{exec}\NormalTok{(}\StringTok{"LRANGE mylist 0 1"}\NormalTok{)}
\CommentTok{\#  [[1]]}
\CommentTok{\#  [1] "vanilla"}
\CommentTok{\#  }
\CommentTok{\#  [[2]]}
\CommentTok{\#  [1] "strawberry"}
\end{Highlighting}
\end{Shaded}

The \textbf{RcppRedis} list commands (under the `standard' names) work
on serialized R objects so we once again utilize the \texttt{exec}
command to execute this using the `standard' name. As access to
unserialized data is useful, the package also two alternates for numeric
and string return data:

\begin{Shaded}
\begin{Highlighting}[]
\NormalTok{redis}\SpecialCharTok{$}\FunctionTok{listRangeAsStrings}\NormalTok{(}\StringTok{"mylist"}\NormalTok{, }\DecValTok{0}\NormalTok{, }\DecValTok{1}\NormalTok{)}
\CommentTok{\#  [1] "vanilla"    "strawberry"}
\end{Highlighting}
\end{Shaded}

\hypertarget{data-structure-example-five-sorted-sets}{%
\section{Data Structure Example Five: Sorted
Sets}\label{data-structure-example-five-sorted-sets}}

\href{https://redis.io}{Redis} offers another data structure that can be
of interest to us for use in time series. Recall how packages
\textbf{zoo} \citep{CRAN:zoo} and \textbf{xts} \citep{CRAN:xts} are,
essentially, indexed containers around (numeric) matrices with a sorting
index. This is commonly the \texttt{Date} type in \textsf{R} for daily
data, or a \texttt{POSIXct} datimetime type for intra-daily data at
approximately a microsecond resolution. One can then index by day or
datetime, subset, merge, \ldots{} We can store such data in
\textsf{Redis} using sorted sets using the index as the first column. A
quick \textsf{R} example illustrates.

\begin{Shaded}
\begin{Highlighting}[]
\NormalTok{m1 }\OtherTok{\textless{}{-}} \FunctionTok{matrix}\NormalTok{(}\FunctionTok{c}\NormalTok{(}\DecValTok{100}\NormalTok{, }\DecValTok{1}\NormalTok{, }\DecValTok{2}\NormalTok{, }\DecValTok{3}\NormalTok{), }\DecValTok{1}\NormalTok{)}
\NormalTok{redis}\SpecialCharTok{$}\FunctionTok{zadd}\NormalTok{(}\StringTok{"myz"}\NormalTok{, m1) }\CommentTok{\# add m1 indexed at 100}
\CommentTok{\#  [1] 1}
\NormalTok{m2 }\OtherTok{\textless{}{-}} \FunctionTok{matrix}\NormalTok{(}\FunctionTok{c}\NormalTok{(}\DecValTok{105}\NormalTok{, }\DecValTok{2}\NormalTok{, }\DecValTok{2}\NormalTok{, }\DecValTok{4}\NormalTok{), }\DecValTok{1}\NormalTok{)}
\NormalTok{redis}\SpecialCharTok{$}\FunctionTok{zadd}\NormalTok{(}\StringTok{"myz"}\NormalTok{, m2) }\CommentTok{\# add m1 indexed at 105}
\CommentTok{\#  [1] 1}
\end{Highlighting}
\end{Shaded}

In this first example we insert two matrices (with three values each)
index at 100 and 105, respectively, to the sorted set under key
\texttt{myz}. We will then ask for a range of data over the range from
90 to 120 which will include both sets of observations.

\begin{Shaded}
\begin{Highlighting}[]
\NormalTok{res }\OtherTok{\textless{}{-}}\NormalTok{ redis}\SpecialCharTok{$}\FunctionTok{zrangebyscore}\NormalTok{(}\StringTok{"myz"}\NormalTok{, }\DecValTok{90}\NormalTok{, }\DecValTok{120}\NormalTok{)}
\NormalTok{res}
\CommentTok{\#       [,1] [,2] [,3] [,4]}
\CommentTok{\#  [1,]  100    1    2    3}
\CommentTok{\#  [2,]  105    2    2    4}
\end{Highlighting}
\end{Shaded}

\hypertarget{communication-example-publishsubscribe}{%
\section{Communication Example:
Publish/Subscribe}\label{communication-example-publishsubscribe}}

We have seen above that writen a \emph{value} to a particular \emph{key}
into a list, set, or sorted set is straightforward. So is publishing
into a \emph{channel}. \textsf{Redis} keeps track of the current
subscribers to a channel and dispatches the published content.

Subscribers can join, and leave, anytime. Data is accessible via the
publish/subscribe (or ``pub/sub'') mechanism while being subscribe.
There is no mechanism \emph{within pub/sub} to obtain `older' values, or
to re-request values. Such services can however be provided by
\textsf{Redis} is its capacity of a data store.

As subscription is typically blocking, we cannnot show a simple example
in the vignette. But an illustration (without running code) follows.

\begin{Shaded}
\begin{Highlighting}[]
\NormalTok{ch1 }\OtherTok{\textless{}{-}} \ControlFlowTok{function}\NormalTok{(x) \{ }\FunctionTok{cat}\NormalTok{(}\StringTok{"Got"}\NormalTok{, x, }\StringTok{"in ch1}\SpecialCharTok{\textbackslash{}n}\StringTok{"}\NormalTok{) \}}
\NormalTok{redis}\SpecialCharTok{$}\FunctionTok{subscribe}\NormalTok{(}\StringTok{"ch1"}\NormalTok{)}
\end{Highlighting}
\end{Shaded}

Here we declare a callback function which by our convention uses the
same name as the channel. So in the next when the subscription is
activated, the callback function is registered with the current
\textsf{Redis} object. Once another process or entity publishes to the
channel, the callback function will be invoked along with the value
published on the channel.

\hypertarget{application-example-hash-r-objects}{%
\section{Application Example: Hash R
Objects}\label{application-example-hash-r-objects}}

The ability to serialize R object makes it particularly to store R
objects directly. This can be useful for data sets, and well as
generated data

\begin{Shaded}
\begin{Highlighting}[]
\NormalTok{fit }\OtherTok{\textless{}{-}} \FunctionTok{lm}\NormalTok{(Volume }\SpecialCharTok{\textasciitilde{}}\NormalTok{ . }\SpecialCharTok{{-}} \DecValTok{1}\NormalTok{, }\AttributeTok{data=}\NormalTok{trees)}
\NormalTok{redis}\SpecialCharTok{$}\FunctionTok{hset}\NormalTok{(}\StringTok{"myhash"}\NormalTok{, }\StringTok{"data"}\NormalTok{, trees)}
\CommentTok{\#  [1] 0}
\NormalTok{redis}\SpecialCharTok{$}\FunctionTok{hset}\NormalTok{(}\StringTok{"myhash"}\NormalTok{, }\StringTok{"fit"}\NormalTok{, fit)}
\CommentTok{\#  [1] 0}
\NormalTok{fit2 }\OtherTok{\textless{}{-}}\NormalTok{ redis}\SpecialCharTok{$}\FunctionTok{hget}\NormalTok{(}\StringTok{"myhash"}\NormalTok{, }\StringTok{"fit"}\NormalTok{)}
\FunctionTok{all.equal}\NormalTok{(fit, fit2)}
\CommentTok{\#  [1] TRUE}
\end{Highlighting}
\end{Shaded}

The retrieved model fit is equal to the one we stored in
\href{https://redis.io}{Redis}. We can also re-fit on the retrieved data
and obtain the same coefficient. (The fit object stores information
about the data set which differs here for technical reason internal to
\textsf{R}; the values are the same.)

\begin{Shaded}
\begin{Highlighting}[]
\NormalTok{data2 }\OtherTok{\textless{}{-}}\NormalTok{ redis}\SpecialCharTok{$}\FunctionTok{hget}\NormalTok{(}\StringTok{"myhash"}\NormalTok{, }\StringTok{"data"}\NormalTok{)}
\NormalTok{fit2 }\OtherTok{\textless{}{-}}\NormalTok{ redis}\SpecialCharTok{$}\FunctionTok{hget}\NormalTok{(}\StringTok{"myhash"}\NormalTok{, }\StringTok{"fit"}\NormalTok{)}
\NormalTok{fit3 }\OtherTok{\textless{}{-}} \FunctionTok{lm}\NormalTok{(Volume }\SpecialCharTok{\textasciitilde{}}\NormalTok{ . }\SpecialCharTok{{-}} \DecValTok{1}\NormalTok{, }\AttributeTok{data=}\NormalTok{data2)}
\FunctionTok{all.equal}\NormalTok{(}\FunctionTok{coef}\NormalTok{(fit2), }\FunctionTok{coef}\NormalTok{(fit3))}
\CommentTok{\#  [1] TRUE}
\end{Highlighting}
\end{Shaded}

\hypertarget{summary}{%
\section{Summary}\label{summary}}

This vignettet introduces the \href{https://redis.io}{Redis} data
structure engine, and demonstrates how reading and writing different
data types from different programming languages including \textsf{R},
\textsf{Python} and shell is concise and effective. A final example of
storing an \textsf{R} dataset and model fit further illustrates the
versatility of \textsf{Redis}.


\bibliography{redis}

\begin{thebibliography}{7}
\newcommand{\enquote}[1]{``#1''}
\providecommand{\natexlab}[1]{#1}
\providecommand{\url}[1]{\texttt{#1}}
\providecommand{\urlprefix}{URL }
\expandafter\ifx\csname urlstyle\endcsname\relax
  \providecommand{\doi}[1]{doi:\discretionary{}{}{}#1}\else
  \providecommand{\doi}{doi:\discretionary{}{}{}\begingroup
  \urlstyle{rm}\Url}\fi
\providecommand{\eprint}[2][]{\url{#2}}

\bibitem[{Eddelbuettel and Balamuta(2018)}]{TAS:Rcpp}
Eddelbuettel D, Balamuta JJ (2018).
\newblock \enquote{Extending R with C++: A Brief Introduction to Rcpp.}
\newblock \emph{The American Statistician}, \textbf{72}(1).
\newblock \doi{10.1080/00031305.2017.1375990}.

\bibitem[{Eddelbuettel \emph{et~al.}(2022)Eddelbuettel, Fran\c{c}ois, Allaire,
  Ushey, Kou, Russel, Chambers, and Bates}]{CRAN:Rcpp}
Eddelbuettel D, Fran\c{c}ois R, Allaire J, Ushey K, Kou Q, Russel N, Chambers
  J, Bates D (2022).
\newblock \emph{{Rcpp}: Seamless {R} and {C++} Integration}.
\newblock R package version 1.0.8,
  \urlprefix\url{https://CRAN.R-Project.org/package=Rcpp}.

\bibitem[{Eddelbuettel and Lewis(2022)}]{CRAN:RcppRedis}
Eddelbuettel D, Lewis BW (2022).
\newblock \emph{RcppRedis: 'Rcpp' Bindings for 'Redis' using the 'hiredis'
  Library}.
\newblock R package version 0.2.0,
  \urlprefix\url{https://CRAN.R-Project.org/package=RcppRedis}.

\bibitem[{Ryan and Ulrich(2020)}]{CRAN:xts}
Ryan JA, Ulrich JM (2020).
\newblock \emph{xts: eXtensible Time Series}.
\newblock R package version 0.12.1,
  \urlprefix\url{https://CRAN.R-project.org/package=xts}.

\bibitem[{Sanfilippo(2009)}]{Redis}
Sanfilippo S (2009).
\newblock \enquote{Redis In-memory Data Structure Server.}
\newblock \url{https://redis.io}.

\bibitem[{Seguin(2012)}]{Seguin:2021:Redis}
Seguin K (2012).
\newblock \enquote{{The Little Redis Book}.}
\newblock \url{https://www.openmymind.net/redis.pdf}.

\bibitem[{Zeileis \emph{et~al.}(2021)Zeileis, Grothendieck, and
  Ryan}]{CRAN:zoo}
Zeileis A, Grothendieck G, Ryan JA (2021).
\newblock \emph{zoo: S3 Infrastructure for Regular and Irregular Time Series
  (Z's Ordered Observations)}.
\newblock R package version 1.8-9,
  \urlprefix\url{https://CRAN.R-project.org/package=zoo}.

\end{thebibliography}
\bibliographystyle{jss}

\end{document}